\documentclass[twocolumn,showpacs,preprintnumbers,amsmath,amssymb,letterpaper]{revtex4}

\usepackage{graphicx}
\usepackage{dcolumn}
\usepackage{bm}
\usepackage{times}

\newcommand{\Tc}{$T_{c}$}
\newcommand{\Tb}{TbTe$_3$}
\newcommand{\R}{$R$Te$_3$}
\newcommand{\Tone}{$T_{CDW}$}
\newcommand{\Ttwo}{$T^{\ast}$}

\begin{document}

\preprint{preprint}

\title{Pressure induced superconductivity in the charge density wave compound terbium tritelluride}

\author{J.\ J.\ Hamlin, D.\ A.\ Zocco, T.\ A.\ Sayles, and M.\ B.\ Maple}
\affiliation{Department of Physics and Institute for Pure and Applied Physical Sciences,\\ University of California, San Diego, La Jolla, CA 92093}
\author{J.\ -H.\ Chu and I.\ R.\ Fisher}
\affiliation{Department of Applied Physics, Geballe Laboratory for Advanced Materials, Stanford University, CA 94305}

\date{\today}

\begin{abstract}
	A series of high-pressure electrical resistivity measurements on single crystals of TbTe$_{3}$ reveal a complex phase diagram involving the interplay of superconducting, antiferromagnetic and charge density wave order.  The onset of superconductivity reaches a maximum of almost 4 K (onset) near $\sim 12.4$ GPa.
\end{abstract}


\maketitle

It has long been known that layered compounds with a high degree of structural anisotropy offer a promising avenue in the search for materials with high superconducting critical temperatures $T_c$ \cite{ginzburg_book_1}.  The highest temperature superconductors known today, the copper-oxide based high-$T_c$ materials, are strongly layered.  Very recently, the pace of the search for layered compounds possesing high superconducting critical temperatures has greatly accelerated following the discovery of superconductivity (SC) at temperatures as high as 55 K in a class of layered materials containing FeAs planes \cite{kamihara_2008_1,ren_2008_1}.  Interestingly, the parent compounds of the cuprate and FeAs-based high temperature superconductors display antiferromagnetic Mott insulating and spin density wave instabilities, respectively, and only become superconducting when the instability is suppressed towards zero temperature through pressure or doping.  Indeed, over the past several decades it has become clear that SC very often appears when a second order phase transition is driven towards zero temperature near a so-called quantum critical point.

The quasi-2D rare-earth tritelluride compounds \R\ ($R$ = La-Nd, Sm, and Gd-Tm) have lately received significant attention as the first system in which nominal square-planar symmetry is broken by the formation of a unidirectional charge density wave (CDW) \cite{kapitulnik_2007_1}.  Furthermore, the CDW transition temperatures of these compounds display striking systematics across the rare-earth series \cite{dimasi_1995_1,fisher_2006_1,ru_2008_1,fisher_2008_1}.  They crystallize in a weakly orthorhombic $Cmcm$ structure composed of double layers of planar Te sheets separated by corrugated $R$Te layers.  For this space group, the crystalline $b$-axis is perpendicular to the Te sheets.  Because large areas of Fermi surface are parallel and may be connected by a single nesting vector, these compounds are unstable to the formation of an incommensurate CDW within the $ac$-planes.  For tritellurides containing the heavier rare-earths Dy-Tm, a second charge density wave, orthogonal to the first, forms at lower temperatures \cite{ru_2008_1,fisher_2008_1}.  These compounds also display magnetic order at or below $\sim 10$ K \cite{suzuki_2003_1,fisher_2008_1}.  Electronic structure calculations and ARPES measurements indicate that the the $R$Te layers play little or no role in the Fermi surface which is instead determined by the Te sheets \cite{suzuki_2004_1,fisher_2005_1,fisher_2008_2}.  Therefore, rare-earth substitution may be understood as applying ``chemical pressure'' in that it primarily alters the lattice parameters without effecting the band filling or structure type \cite{dimasi_1995_1}.  It has been found that the CDW transition temperatures of the \R\ compounds correlate remarkably well with the in-plane lattice parameters.  A reduction in lattice parameter suppresses the upper CDW transition temperature and enhances the lower CDW transition temperature \cite{ru_2008_1}.

Sacchetti \textit{et al.}, \cite{degiorgi_2007_1,fisher_2006_1}, using optical reflectivity, found that the upper CDW gap of CeTe$_3$ closes under increasing external pressure, leading to the conclusion that external and chemical pressure have qualitatively similar effects.  However, to date there appear to have been no high-pressure transport studies of rare-earth tritellurides.  Therefore, we undertook to study the influence of external pressure on the various ordering temperatures of \Tb\ and to investigate the possibility that SC might appear if any of these ordering temperatures could be driven towards zero temperature.  At ambient pressure, \Tb\ orders magnetically at $\sim 6$ K and displays CDW ordering near 340 K \cite{suzuki_2003_1,fisher_2008_1}.  A second orthogonal CDW has not been observed in \Tb, although previous data suggests \cite{fisher_2008_1} that one might be induced by a small reduction in lattice parameter.

In this Letter, we report the results of ambient pressure calorimetric and high-pressure resistivity measurements on single crystalline \Tb.  The specific heat measurements show that, at ambient pressure, \Tb\ is non-superconducting down to 600 mK.  Under pressure, the upper CDW transition temperature is suppressed.  A second feature in the resistivity, appearing above 1.2 GPa and moving to higher temperatures with pressure, is consistent with the appearance of a second, lower temperature, CDW.  Above 2.3 GPa, we find that \Tb\ becomes superconducting with \Tc\ (onset) reaching nearly 4 K near $\sim 12.4$ GPa.  Remarkably, at 2.3 GPa, three types of order, charge density wave, antiferromagnetism, and SC all appear upon progressively cooling the sample.

Single crystals of TbTe$_3$ were grown by slow cooling of a binary melt as described elsewhere \cite{fisher_2006_2}.  Specific heat $C(T)$ measurements were made $\sim 0.6-9$ K in a home-built $^3$He calorimeter using a semi-adiabatic heat-pulse technique. Single crystals were attached to a sapphire platform with a small amount of Apiezon N grease.  For the high-field calorimetric measurements, the magnetic field was applied parallel to the crystallographic $b$-axis.

We performed three high-pressure resistivity experiments on TbTe$_3$.  The first experiment, performed in a Teflon capsule piston-cylinder cell, reached a maximum pressure of 2.3 GPa and utilized a nearly hydrostatic, 1:1 mixture of n-pentane and isoamyl alcohol as the pressure transmitting medium.  Two other experiments, performed in a Bridgman anvil cell, reached maximum pressures of 15.2 GPa and 12.4 GPa, respectively.  The Bridgman anvil cell experiments utilize quasi-hydrostatic, solid steatite as the pressure transmitting medium, tungsten-carbide anvils with 3.5 mm diameter flats and a total of eight wires to measure the resistance of samples of \Tb, CeTe$_3$, and a Pb manometer.  The results of the CeTe$_3$ experiments will be reported in a separate paper.  In all of the experiments, pressure was determined from the superconducting transition temperature of a strip of Pb foil using the calibration of Bireckoven and Wittig \cite{wittig_1988_1}.  The magnitude of pressure gradients over the sample can be estimated from the width of the Pb superconducting transition.  Pressure gradients were as large as $\pm 2 \%$ of the total pressure for the piston-cylinder experiment and $\pm 10 \%$ for the Bridgman anvil cell experiment.  The resistance of each sample was measured in the \textit{ac}-plane using 4 wires and a Linear Research Inc.\ LR-700 AC resistance bridge.  Due to uncertainties in the geometry of the small samples and placement of the leads (which may move during pressurization), the absolute resistivity values are only accurate to within a factor of $\sim 4$.

\begin{figure}
	\begin{center}
		\includegraphics[width=\columnwidth]{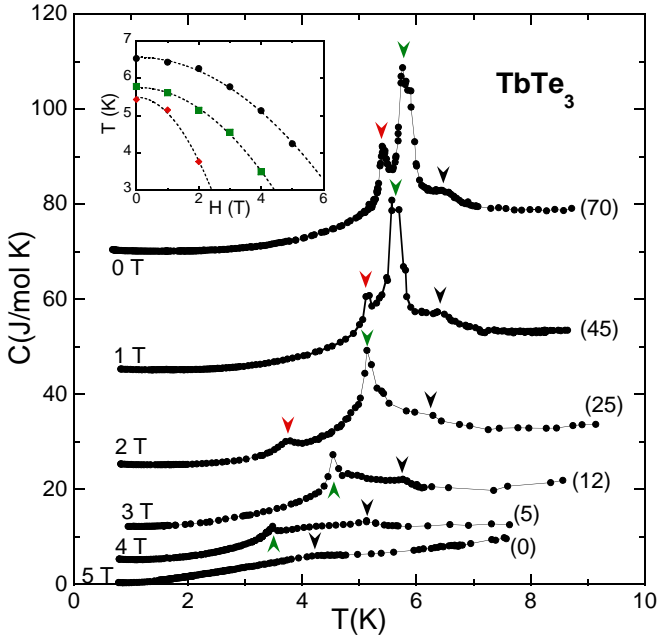}
	\end{center}
	\caption{(color online only) Specific heat $C$ versus temperature $T$ for TbTe$_3$ at ambient pressure and fields up to 5 T.  For clarity, the curves have been offset vertically by a constant amount indicated in parentheses near the right side of the curves.  Three distinct peaks, indicated by arrows, are visible near the N\'{e}el temperature at 6 K. (\textit{inset})  The field dependence of the specific heat anomalies.  The dashed line fits are described in the text.}
	\label{fig:fig1}
\end{figure}
In figure~\ref{fig:fig1}, we plot the results of our specific heat measurements.  The importance of these results is twofold: they lend further weight to the notion that \Tb\ orders antiferromagnetically at low temperature and they establish the lack of superconductivity in the previously uninvestigated temperature range 600 mK - 2 K.  Similar to recently reported results \cite{fisher_2008_1}, we observe several peaks near $T_N$.  A weak upturn at the lowest temperatures is most likely due to a nuclear Schottky anomaly.  Under applied magnetic fields, all three of the peaks become smaller and move down in temperature with increasing field at differing rates.  The inset to figure~\ref{fig:fig1} shows the field dependence of these peaks.  In a molecular field approximation analysis of the magnetic phase diagram of an antiferromagnet, the ordering temperature $T$ is expected to depend on the applied field $H$ as $T=T_N-b\cdot (H^2/T_N)$, where $b$ is a constant \cite{foner_1970_1}.  The dashed lines shown in the lower inset to figure~\ref{fig:fig1} show that the data are well fit by the above equation affirming that all three peaks are likely due to a series of closely spaced antiferromagnetic transitions.  Extrapolation of the fits implies that these transitions would be suppressed to zero temperature at 8.6, 6.4 and 3.6 tesla, respectively.

\begin{figure}
	\begin{center}
		\includegraphics[width=\columnwidth]{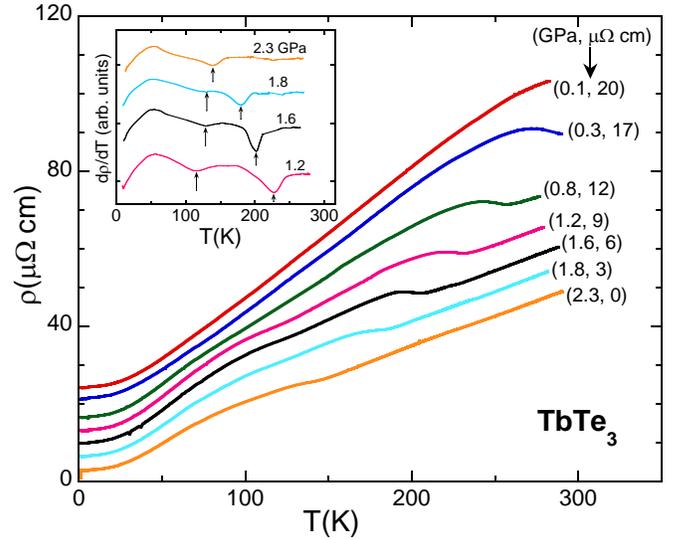}
	\end{center}
	\caption{(color online only) Electrical resistivity measured in the \textit{ac}-plane of TbTe$_3$ under nearly-hydrostatic pressures to 2.3 GPa.  The curves have been offset vertically for clarity.  The pressure and magnitude of the offset are indicated in parentheses at the right of the curves.  The onset of the CDW is clearly seen to shift downward with pressure.  At and above 1.2 GPa, a second feature in the resistivity is clearly visible as a minimum in the slope $d\rho /dT$ (\textit{inset}).  At 2.3 GPa the two transitions \Tone\ and \Ttwo\ appear to be nearly coincident.}
	\label{fig:fig2}
\end{figure}
Figure \ref{fig:fig2} shows the result of the nearly-hydrostatic piston-cylinder cell high-pressure measurements on \Tb.  The onset of CDW ordering, \Tone, first becomes visible at 0.3 GPa as a kink in the resistivity just below room temperature.  Upon further increasing the pressure, \Tone\ decreases monotonically to $\sim 140$ K at 2.3 GPa.  A measurement taken on unloading from 2.3 to 1.2 GPa indicates that the pressure dependence of \Tone\ is reversible.  A second feature, \Ttwo, below \Tone\ is visible as a minimum in the temperature derivative of the resistivity $d\rho /dT$ (inset to figure~\ref{fig:fig2}).  Under pressure, \Ttwo\ moves to higher temperatures until, at 2.3 GPa, \Tone\ and \Ttwo\ can no longer be distinguished.  We surmise that the feature at \Ttwo\ is related to the appearance of a second CDW, orthogonal to the first, as observed at ambient pressure for the heavy rare earth tritellurides \cite{fisher_2008_1}, although this will have to be confirmed through direct measurements of the electron distribution under pressure.  Also at 2.3 GPa, a sharp drop in the resistivity, suggestive of SC, appeared near 2 K.

\begin{figure}
	\begin{center}
		\includegraphics[width=\columnwidth]{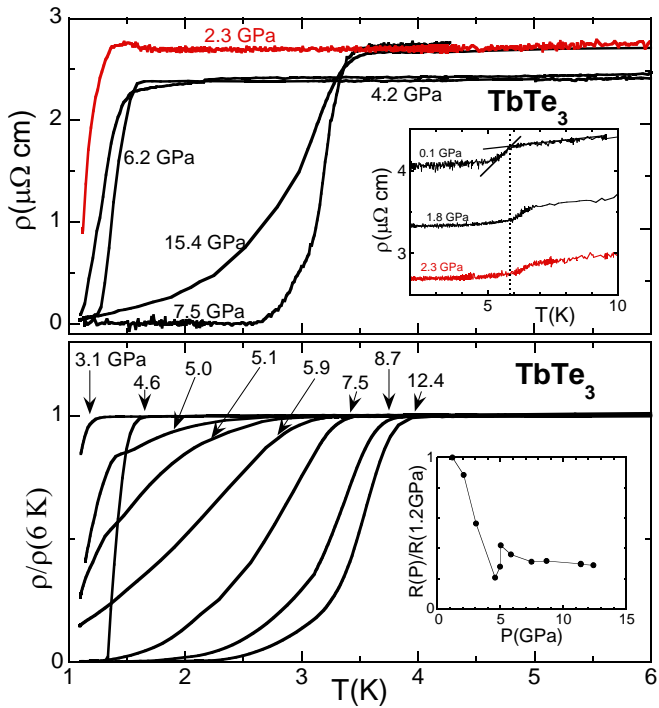}
	\end{center}
	\caption{(color online only) Electrical resistivity measured in the \textit{ac}-plane of TbTe$_3$ at various pressures.  The upper panel shows data from the piston-cylinder cell and first Bridgman anvil cell experiment while the lower panel shows data taken from the second Bridgman anvil cell experiment.  At several pressures, the resistance becomes negligible below the transition confirming SC.  \textit{upper inset:} Resistivity versus temperature in the vicinity of $T_N$ at selected pressures.  The dashed and solid black lines in the inset indicate the value of $T_N$ at 0.1 GPa. \textit{lower inset:} Relative change in the 6 K resistance, taken from the second Bridgman anvil cell experiment.}
	\label{fig:fig3}
\end{figure}
In order to further investigate the possibility of pressure induced SC in \Tb, we performed additional high-pressure experiments in a Bridgman anvil cell.  In these measurements we were unable to track \Tone, \Ttwo, or $T_N$ to higher pressures because appreciable pressure gradients in the Bridgman anvil cell lead to a broadening of the already weak resistive anomalies so that the transitions could not be unambiguously pinpointed.  However, the Bridgman anvil cell data did allow us to confirm and further explore the superconducting state.  Figure~\ref{fig:fig3} illustrates the low temperature behavior of the resistivity of \Tb\ from the piston-cylinder and first Bridgman anvil cell experiment (upper panel) and second Bridgman anvil cell experiment (lower panel).  Several measurements display complete resistive transitions to an immeasurably small resistivity below \Tc, providing clear evidence that \Tb\ becomes superconducting under pressure.  The onset \Tc\ reaches nearly 4 K.  The upper inset of figure~\ref{fig:fig3} shows the resistivity versus temperature in the vicinity of $T_N$ at several pressures measured in the piston-cylinder cell.  The N\'{e}el temperature increases monotonically with pressure over the measured pressure range.  At 2.3 GPa (red curves in figure~\ref{fig:fig3}), the antiferromagnetic (AFM) resistive anomaly and SC are both present.  Near a maximimum in \Tc\ versus pressure, pressure gradients have a minimal effect on the spread of \Tc\ values and thus the evolution of the width of the SC transition in our data suggests that \Tc\ may pass through a maximum near 12-15 GPa.

\begin{figure}
	\begin{center}
		\includegraphics[width=\columnwidth]{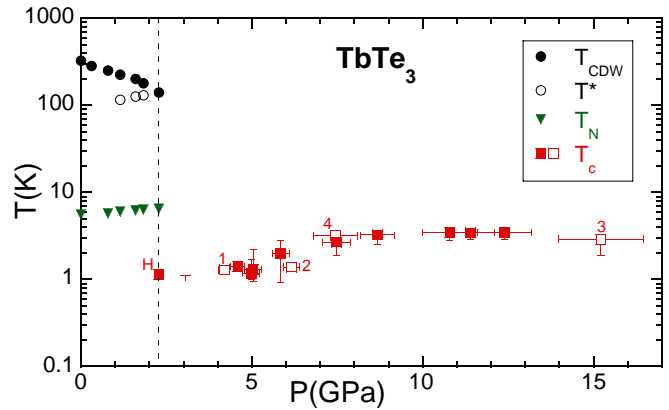}
	\end{center}
	\caption{(color online only) Pressure dependence of the transition temperatures of the phases observed in TbTe$_3$.  The value of $T_c$ is given by the transition midpoint and the vertical error bars are determined by the 10\%-90\% width of the transition (the 10\% value has been estimated via extrapolation where necessary).  The point labeled 'H' is from the hydrostatic cell run.  The numbers indicate the order of measurement for the first Bridgman anvil cell run.  Open squares are from the first Bridgman anvil cell run.}
	\label{fig:fig4}
\end{figure}

Figure~\ref{fig:fig4} presents a summary of all of our our data in the form of a pressure-temperature phase diagram.  The vertical dashed line near 2.3 GPa indicates the upper pressure limit for our hydrostatic piston-cylinder cell experiments; points beyond this line were measured using the Bridgman anvil cell technique.  The lack of \Tone, \Ttwo, and $T_N$ data points at pressures above this dashed line is due to our inability to resolve these transitions in the Bridgman anvil cell, as discussed above, and does not necessarily indicate that the transitions are completely suppressed in this pressure region.  We are investigating the possibility of tracking these transitions to higher pressures through the use of a more hydrostatic pressure medium or high-pressure ac calorimetric measurements.  We find no evidence for SC at ambient pressure and pressures below 2.3 GPa down to 600 mK and $\sim 1.1$ K, respectively. At 2.3 GPa, CDW, AFM and SC order all appear upon cooling the sample.

Our measurements suggest that \Tb\ under pressure is likely a magnetically ordered superconductor in which long range AFM order coexists with SC.  It would be illuminating to follow $T_N$ to higher pressures to determine its evolution and relation to SC. If $T_N$ drops below $T_c$, measurements in the superconducting state could reveal $T_N$ through, for example, measurements of the upper critical field $H_{c2}$ versus temperature which can be either enhanced or depressed below the N\'{e}el temperature \cite{maple_1982_1}, or through features in the specific heat.  A systematic study of the remaining rare-earth tritellurides under pressure ought to be performed in order to determine whether they also display SC and if the $T_c$ values follow a de Gennes' scaling.  If they do, then \Tb\ would be expected to posses amongst the lowest $T_c$ values in the series and significantly higher $T_c$ values would be expected in, for example, LaTe$_3$ under pressure.

Though a purely electronic origin is the simplest explanation for the existence of the two CDWs at ambient pressure in Tm, Er, Ho and DyTe$_3$ \cite{ru_2008_1}, it is conceivable that the progressive change in mass on going across the rare-earth series is somehow playing a role. However, if, as we propose, \Ttwo\ proves to be related to the formation of a second CDW in \Tb, our results lend substantial weight to the arguments based on the changing band-width as the origin of the second CDW.  It would be of particular interest to study \Tone\ and \Ttwo\ to higher pressures in order to determine if they are suppressed near the same pressure at which SC appears or reaches a maximum \Tc.  It is possible that the SC in \Tb\ may be understood in terms of a Bilbro-McMillan partial gaping scenario \cite{mcmillan_1976_1} in which the SC and CDW compete for the Fermi surface and, when the CDW is suppressed, SC is enhanced as additional Fermi surface becomes available to the superconducting state.  The weak pressure dependence of \Tc\ beyond 7.5 GPa makes it likely that the superconductivity does not derive from quantum critical fluctuations, but rather from Fermi surface competition.

In summary, we have determined the effect of pressure on the CDW and magnetic ordering temperatures of \Tb\ and, to our knowledge, observed the first example of SC in a rare-earth tritelluride.  Superconductivity first appears in \Tb\ at 2.3 GPa near 1.2 K and reaches a maximum of $\sim 4$ K (onset) near 12.4 GPa.  It appears that the rare-earth tritellurides under pressure may offer an ideal class of compounds for the systematic study of the interplay and coexistence of charge density waves, magnetic order and superconductivity.
\vspace{-5mm}
\begin{acknowledgments}
\vspace{-3mm}
	Research at the University of California, San Diego, was supported by the U.\ S.\ Department of Energy (DOE) grant number DE-FG52-06NA26205.  Crystal growth at Stanford University was supported by the U.\ S.\ DOE under contract No.\ DE-AC02-76SF00515.
\end{acknowledgments}

\end{document}